\documentclass[10pt,conference]{IEEEtran}
\IEEEoverridecommandlockouts
\usepackage{amsmath,amssymb,amsthm,bm,mathtools}
\usepackage{graphicx}
\usepackage{enumitem}
\usepackage{mathrsfs}
\usepackage{algorithm}
\usepackage{booktabs}
\usepackage{multirow}
\usepackage[caption=false,font=footnotesize]{subfig}

\newtheorem{proposition}{Proposition}
\newtheorem{theorem}{Theorem}

\begin{document}

	\title{Movable Antenna-Enabled Near-Field Localization: A Moment-Based Approach}
	\author{
		\IEEEauthorblockN{Jinjian Liu,
			Xianxin Song, and
			Xianghao Yu}
		\IEEEauthorblockA{Department of Electrical Engineering, City University of Hong Kong, Hong Kong}
		\IEEEauthorblockA{Emails: jinjian.liu@my.cityu.edu.hk, xianxin.song@cityu.edu.hk, alex.yu@cityu.edu.hk}
	}
	\maketitle
	\begin{abstract}
		Movable antennas (MAs) have recently emerged as a promising technology to enhance the wireless sensing performance by exploiting additional spatial degrees of freedom. In this work, we consider an MA-enabled near-field target localization problem under a device-based sensing architecture, where the sensing target actively transmits signals and the sensing receiver estimates its location based on the received signals. We adopt the squared position error bound (SPEB) as the sensing performance metric and aim to minimize the target's SPEB by optimizing the antenna placement at the sensing receiver. The resulting optimization problem is highly non-convex due to the minimum antenna spacing constraints and the intricate dependence of the SPEB on the antenna locations  and the intractable form of SPEB. To tackle this problem, we first relax the inter-spacing constraints and accordingly formulate an antenna position distribution problem in terms of degree-4 truncated moments. The reformulated moment-based problem admits an equivalent semidefinite programming formulation and can be optimally solved using standard convex optimization solvers. Using truncated moment theory, we then recover an optimal antenna position distribution supported on a finite number of locations and construct an efficient clustered placement policy. Numerical results demonstrate that the constructed optimal distribution exhibits a three-cluster structure with source-dependent weights. The proposed antenna placement method outperforms conventional array geometries and closely approaches the exhaustive-search benchmark scheme.
	\end{abstract}
\vspace{-1mm}
\section{Introduction}

Driven by the rapidly growing demand for wireless sensing, movable antennas (MAs) have recently emerged as a promising technology for next-generation sensing systems \cite{MaTWC0824}. 
Unlike conventional fixed-position antennas, MAs allow antenna positions to be flexibly adjusted within given regions, providing additional spatial degrees of freedom associated with antenna positions to enhance wireless sensing performance.  
To fully exploit its potential, MA systems typically employ large aperture array in practice \cite{ZhuWCNC}, while it also expands the near-field region, rendering the conventional plane-wave approximation inaccurate. 
Accordingly, a spherical-wave model is adopted for near-field propagation \cite{liuOJCS2023} to capture the wavefront curvature and exploit both the angle and range information of the source for target localization.

Recently, the antenna position policies of MA have been extensively investigated in both the near-field wireless sensing \cite{wangicc2025, liuarxiv2609} and integrated sensing and communications (ISAC) \cite{DingTWC2026, perović2025arxiv}. For example, the authors in \cite{wangicc2025} considered a device-based single-target sensing scenario, in which an MA array receives signals emitted by the sensing target, aiming to minimize the worst-case sum Cram\'{e}r-Rao bound (CRB) for angle and range estimation by optimizing antenna positions via an iterative algorithm. The authors in \cite{liuarxiv2609} considered a monostatic single-target wireless sensing system, in which colocated transmit and receive arrays actively transmit sensing signals toward the target and collect the reflected echoes, respectively. To minimize the worst squared position error bound (SPEB) over the entire near-field region, they established the optimality of centro-symmetric antenna position distributions, and accordingly derived a closed-form solution for the joint optimization of transceiver antenna placements and transmit power allocation. 
The authors in \cite{DingTWC2026} investigated a monostatic ISAC system, where a base station uses transmit/receive MA arrays to emit signals and collect echoes from multiple targets while serving uplink and downlink users. They jointly optimized the beamformers, sensing signal covariance matrices, antenna positions, and uplink power allocation to maximize the weighted sum of sensing and communication rates.
The authors in \cite{perović2025arxiv} considered a multi-user, single-target ISAC system, where a base station employs a transmit MA array to serve MA-equipped users and transmit signals toward the target, while a fixed receive array collects the target echoes. They proposed an iterative algorithm to jointly optimize the MA positions, communication precoders, sensing transmit beamformer, and receive combiner to maximize the weighted sum rate of the communication users, subject to a minimum sensing signal-to-interference-plus-noise ratio constraint.

However, existing studies on MA-enabled near-field sensing \cite{wangicc2025} and ISAC \cite{DingTWC2026, perović2025arxiv} primarily rely on complex iterative optimization, which leads to high computational complexity when the number of antennas is large and provides limited theoretical insight into optimal antenna placement policies.
Additionally, prior studies \cite{wangicc2025, liuarxiv2609} adopted the worst-case estimation performance over the entire near-field region as the sensing performance metric. This robustness-oriented design can effectively mitigate uncertainty on the sensing target's location. However, it may lead to degraded sensing performance when the location of sensing target can be accurately predicted, as the optimal antenna placement structure generally depends on the specific sensing target location.
This motivates location-specific antenna placement design, particularly in target tracking scenarios where the predicted target position can be leveraged to optimize the array geometry.

In this paper, we investigate the antenna placement for an arbitrary target's position in a device-based MA-enabled near-field wireless sensing system. 
We first establish the near-field system model and adopt SPEB as the sensing performance metric. By temporarily relaxing the inter-spacing constraint, we then reformulate the antenna placement problem in terms of the degree-4 truncated moments of the antenna position distribution. 
The resulting moment-based problem is shown to be convex and admits an equivalent semidefinite programming (SDP) formulation, yielding an optimal  moment sequence. Based on this result, we subsequently recover an optimal antenna position distribution concentrated at a finite number of locations, leading to a clustered antenna deployment strategy.
 Numerical results demonstrate that the proposed design outperforms conventional benchmark schemes and achieves performance close to that of the exhaustive search method, thereby validating its effectiveness.

\textit{Notations:}
We use normal-face letters to denote scalars,
boldface lowercase and uppercase letters to denote column vectors and matrices, respectively.
The imaginary unit is denoted by $\jmath = \sqrt{-1}$.
The nearest-integer rounding operator is denoted by
$\operatorname{round}(\cdot)$.
For a vector \(\mathbf z\),
\(\operatorname{diag}(\mathbf z)\) denotes the diagonal matrix with
\(\mathbf z\) on its main diagonal. 
For a matrix $\mathbf Z$, $\operatorname{tr}(\mathbf Z)$ and
$\mathrm{col}_j(\mathbf Z)$ denote its trace and its $(j+1)$-th
column, respectively. The notation $\big[b_{ij}\big]_{0\le i,j\le k}$ denotes
the $(k+1)\times(k+1)$ matrix whose $(i+1,j+1)$ entry
is $b_{ij}$.
The transpose, expectation and Loewner partial order operations are denoted by $(\cdot)^{\mathrm{T}}$,  $\mathbb{E}(\cdot)$ and $\preceq$/$\succeq$, respectively.
The $N\times N$ identity matrix is denoted by $\mathbf I_N$.
The set of $M \times N$ real matrices is denoted by $\mathbb{R}^{M \times N}$.
The set of complex numbers is denoted by $\mathbb{C}$. The set of $n\times n$ real symmetric matrices is denoted
by $\mathbb S^n$.
A circularly symmetric complex Gaussian vector with mean $\bm \mu$ and covariance $\mathbf A$
is denoted by $\mathcal{CN}(\bm \mu,\mathbf A)$.
The Dirac point mass at $x$, which assigns probability one to $x$,
is denoted by $\delta_x(\cdot)$.
	\section{System Model and Problem Formulation}
\begin{figure}[t]
	\centering
	\includegraphics[width=0.65\linewidth, height=0.65\linewidth, keepaspectratio]{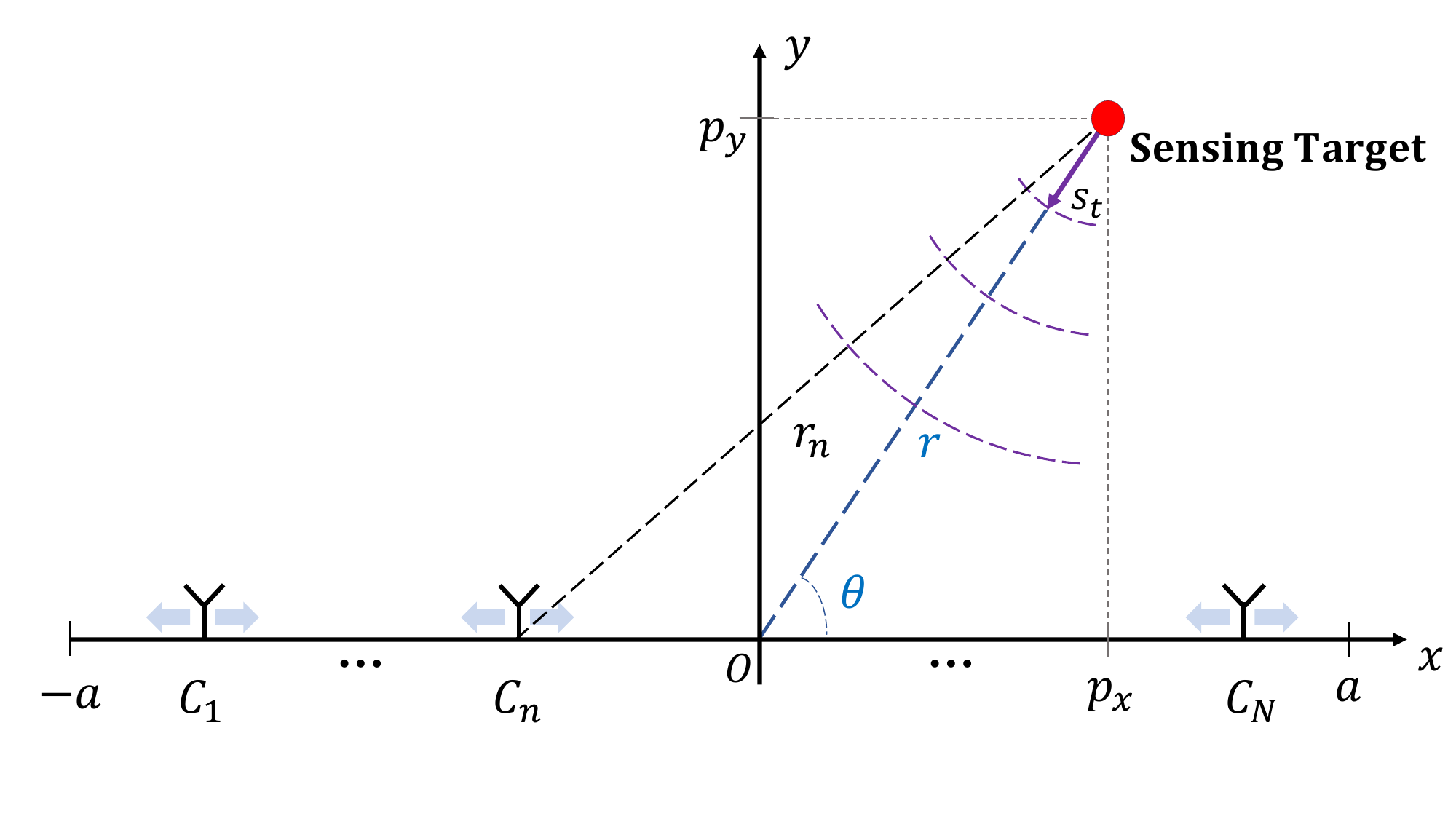}
	\vspace{-6mm}
	\caption{The MA-enhanced near-field sensing system.}
	\vspace{-6mm}
	\label{fig:nearfield_scenario}
\end{figure}

	We consider a single-source near-field sensing system as shown in Fig.~\ref{fig:nearfield_scenario}, consisting of a sensing target located within the near-field region \cite{liuOJCS2023} and a sensing receiver equipped with an MA array. In particular, the sensing target actively emits sensing signals, while the sensing receiver estimates the target's location by processing the received signals.

\subsection{System Model}

First, we introduce the transmit signal model. Let $T$ denote the number of sensing snapshots.
At snapshot $t \in \{1, \dots, T\}$, the sensing target transmits a narrowband deterministic sensing waveform $s_t$ with wavelength $\lambda$ \cite{MaTWC0824}. This waveform is subject to the average power constraint $\frac{1}{T} \sum_{t=1}^{T} |s_t|^2 = P$, with $P$ being the transmit power.

 Next, we present the sensing channel model. The linear MA array consists of $N$ receive antennas deployed along the $x$-axis within a given aperture $[-a, a]$. The position of the $n$-th antenna is given by $C_n = (x_n, 0)$ for $n = 1, \ldots, N$. Without loss of generality, the antennas are indexed from left to right, i.e., $-a \le x_1 < x_2 < \ldots < x_N \le a$, with their coordinates collected in $\mathbf{x} = [x_1, \ldots, x_N]^{\mathrm{T}}$.
Let $\mathbf{p}=[p_x,p_y]^{\mathrm{T}}$ denote the Cartesian coordinates of the sensing target. Accordingly, the angle of arrival and the distance from the array center to the sensing target
are respectively given as $\theta(\mathbf{p}) = \arctan\left(\frac{p_y}{p_x}\right)$ and $r(\mathbf{p}) = \sqrt{p_x^2+p_y^2}$. For notational brevity, we write $\theta$ and $r$ instead of $\theta(\mathbf{p})$ and $r(\mathbf{p})$ hereafter.
 Then, the distance from the source to the $n$-th receive antenna is given by
$
r_n=\sqrt{r^2-2r\cos\theta x_n+x_n^2}.
$
We adopt the uniform spherical wave (USW) channel model over the near-field region $\Gamma$ \cite{liuOJCS2023}, which assumes a uniform path-loss amplitude across the array while capturing the phase variations.
By applying the Fresnel approximation to $r_n$ \cite{liuOJCS2023} and omitting the common range-dependent phase term $e^{\jmath\frac{2\pi}{\lambda}r}$, we obtain the array response \cite{wangicc2025}
\begin{equation}
	\label{eq. steering vector}
	\resizebox{0.99\columnwidth}{!}{$\displaystyle
		\mathbf{a}(\mathbf{x},\mathbf{p}) \approx
		\left[
		e^{\jmath \frac{2\pi}{\lambda}
			\left(\cos\theta x_1-\frac{\sin^2\theta}{2r}x_1^2\right)},
		\dots,
		e^{\jmath \frac{2\pi}{\lambda}
			\left(\cos\theta x_N-\frac{\sin^2\theta}{2r}x_N^2\right)}
		\right]^{\mathrm{T}}.
		$}
\end{equation}
The received signal at snapshot $t$ is accordingly modeled as
\begin{equation}
	\mathbf{y}_t=\alpha\beta(r)\mathbf{a}(\mathbf{x},\mathbf{p})s_t+\mathbf{n}_t,\qquad
	t=1,\ldots,T, \label{eq. receive}
\end{equation}
where $\alpha\in\mathbb{C}$ denotes an unknown complex channel coefficient, $\beta(r)=1/r$ models the distance-dependent path-loss, and $\mathbf{n}_t\sim\mathcal{CN}(\mathbf{0},\sigma^2\mathbf{I}_N)$ is the additive white Gaussian noise (AWGN) at the sensing receiver.
In \eqref{eq. receive}, both the source position parameter $\mathbf{p}$ and the channel coefficient parameter $\alpha$ are unknown, with the former being our estimation objective.

	Then, we introduce the sensing performance metric. To characterize sensing target's position estimation performance, we adopt the SPEB, which provides a lower bound on the sum of the mean squared errors in estimating the Cartesian coordinates $p_x$ and $p_y$ for any unbiased position estimator \cite{Wintit2010}. The SPEB at location $\mathbf{p}$ is given by \cite{liuarxiv2609}	
	\begin{equation}
		\mathrm{SPEB}(\mathbf{x}, \mathbf{p}) = \kappa \frac{r^4J_{2}+r^2\sin^2\theta J_{1}}{\sin^2\theta \left(J_{1}J_{2}- J^2_{3}\right)},
		\label{eq.FIM-monostatic}
	\end{equation}
	where $\kappa
	=\frac{\sigma^2\lambda^2}{8\pi^2TPN|\alpha|^2}$,
	$
	J_1
	=
	m_2+\frac{2\cos\theta}{r}m_3+\frac{\cos^2\theta}{r^2}m_4
	-
	\left(m_1+\frac{\cos\theta}{r}m_2\right)^2
	$, $
	J_2
	=
	\frac{\sin^4\theta}{4r^4}\left(m_4-m_2^2\right),
	\label{eq:Gr2}
	$ and $
	J_3
	=
	\frac{\sin^2\theta}{2r^2}
	\left(
	m_3+\frac{\cos\theta}{r}m_4-m_1m_2-\frac{\cos\theta}{r}m_2^2
	\right)
	$, with $m_k=\frac{1}{N}\sum_{n=1}^N x_n^k$.
	\vspace{0mm}
	\subsection{Problem Formulation}
	 For a given source position $\mathbf{p}$, the antenna placement optimization problem is formulated as:
	\begin{subequations}
		\begin{alignat}{2}
			(\mathrm{P}_1):\quad
			& \underset{\mathbf{x}}{\operatorname{minimize}}
			&\quad & \mathrm{SPEB}(\mathbf{x},\mathbf{p}) \\
			& \hspace{0.1cm} \mathrm{s.t.}
			&      & x_n \in [-a,a],
			\quad n=1,\dots,N, \\
			&
			&      & x_{n+1}-x_n \ge \frac{\lambda}{2},
			\quad n=1,\dots,N-1.
			\label{eq. cons-distri-1}
		\end{alignat}
	\end{subequations}
	where constraint \eqref{eq. cons-distri-1} ensures a minimum inter-element spacing to mitigate the mutual coupling effects between adjacent antennas. Here, the spacing constraint \eqref{eq. cons-distri-1} introduces strong coupling among the antenna positions, combined with the complex expression of the SPEB, making problem $(\mathrm{P}_1)$ highly non-convex and mathematically intractable.
	
	\section{Two-stage Algorithm}
	In this section, we first reformulate the antenna placement problem $(\mathrm P_1)$ to an antenna position distribution optimization problem and develop a two-stage algorithm to solve it. In Stage~I, we obtain an optimal degree-$4$ truncated moment sequence via convex optimization. In Stage~II, we construct a finitely atomic representing distribution by leveraging truncated moment theory based on the optimized moment sequence.
	
	\subsection{Problem Reformulation}
	\label{subsec:reformulation}
	To facilitate the optimization, we relax the inter-spacing constraint \eqref{eq. cons-distri-1} and reformulate the antenna placement problem as an optimization over probability distributions rather than discrete antenna positions \cite{liuarxiv2609}. We introduce a discrete random variable $X$ supported on $K$ to characterize the distribution of antenna positions, with its probability mass function as $w(\cdot)$ on $K$ by $w(\zeta)= \Pr(X=\zeta)$ for all $\zeta\in K$, where $K=\{\zeta_1,\ldots,\zeta_L\}\subseteq[-a,a]$ with $L$ a sufficiently large odd integer. Accordingly, $m_i$ can be considered as the $i$-th order moment of $X$, i.e., $m_i=\mathbb{E}(X^i)$, and we let vector $\mathbf m= [m_0,m_1,\ldots,m_4]^\mathrm{T}$. 
	Consequently, the discrete antenna placement problem can be reformulated as the following distribution optimization problem $(\mathrm{P}_{2})$:
	\begin{subequations}
		\begin{alignat}{2}
			(\mathrm{P}_{2})\quad
			& \operatorname*{minimize}_{
				w(\cdot)}
			&\quad&
			f(\mathbf{m}),
			\\
			& \hspace{0.1cm} \mathrm{s.t.}
			&&
			\sum_{\zeta\in K}w(\zeta)=1,
			\label{eq. P2 cons-distri-1}
			\\
			&&&
			w(\zeta)\geq0,\quad
			\forall\zeta\in K.
			\label{eq. P2 cons-distri-2}
		\end{alignat}
	\end{subequations}
    where $f(\mathbf{m})= \frac{1}{\kappa}\mathrm{SPEB}(\mathbf{x}, \mathbf{p})$.
    Problem $(\mathrm{P}{2})$ depends on $w(\cdot)$ only through the degree-$4$ truncated moment sequence
    $\mathbf{m}$, which motivates the following two-stage optimization procedure. Specifically, in the first stage, we optimize over the feasible degree-$4$ moment sequences and denote the optimal solution by $\mathbf m^\star=(m_0^\star,\ldots,m_4^\star)^{\mathrm T}$. Since $\mathbf m^\star$ consists only of moments up to order four, it may admit multiple representing distributions. Hence, in the second stage we recover a feasible antenna position distribution matching $\mathbf m^\star$, thereby enabling a practical antenna placement implementation.

	\subsection{Stage I: Optimal Moment Sequence via SDP}
	In the first stage of solving problem $(\mathrm{P}_2)$, we determine
	the optimal degree-$4$ truncated moment sequence $\mathbf m^\star$ by solving
	the following finite-dimensional moment-domain problem:
	\begin{equation}
		\mathrm{(P_{2.1})}\qquad
		\operatorname*{minimize}_{\mathbf m\in\mathcal{M}_4}
		\quad f(\mathbf{m}),
		\label{eq:P_m}
	\end{equation}
	where $\mathcal{M}_4$ denotes the feasible set of degree-$4$ truncated moment sequences.
	The constraint $\mathbf m\in\mathcal{M}_4$ ensures that all entries of $\mathbf m$ can be realized as the moments of a random variable $X$.  According to the conclusion of the univariate truncated $K$-moment problem
	\cite{schmüdgen2017}, the constraint $\mathbf m\in\mathcal M_4$ in problem $\mathrm{(P_{2.1})}$ is equivalent to the positive semidefiniteness (PSD) of the second-order \textit{localizing matrix}
	$
		\mathbf L_2(\mathbf m)=
		\begin{bmatrix}
			a^2m_0-m_2 & a^2m_1-m_3\\
			a^2m_1-m_3 & a^2m_2-m_4
		\end{bmatrix}
	$
	and the second-order \textit{moment matrix} 
	$
		\mathbf M_2(\mathbf m)=
		\begin{bmatrix}
			m_0 & m_1 & m_2\\
			m_1 & m_2 & m_3\\
			m_2 & m_3 & m_4
		\end{bmatrix}
	$, 
	together with the constraint $m_0=1$.
	These conditions ensure that $\mathbf m$ admits a 
	probability distribution supported on $[-a,a]$. Consequently, the problem $\mathrm{(P_{2.1})}$ can be equivalently reformulated as
		\begin{subequations}
		\begin{alignat}{2}
			\mathrm{(P_{2.1}')}\quad
			& \operatorname*{minimize}_{
				\mathbf{m}\in\mathbb{R}^5}
			&\quad&
			f(\mathbf{m}),
			\\
			& \hspace{0.1cm} \mathrm{s.t.}
			&&
			 \hspace{-0.9cm} \mathbf L_2(\mathbf m)\succeq\mathbf{0},\quad\mathbf M_2(\mathbf m) \succeq \mathbf{0},
			\label{eq. p2-cons-distri-1}
			\\
			&&& \hspace{-0.9cm} m_0=1.
			\label{eq. p2-cons-distri-2}
		\end{alignat}
	\end{subequations}
		It's worth noting that directly verifying the convexity of problem $\mathrm{(P_{2.1})}$ via Hessian matrix analysis is mathematically tedious. 
		To simplify the analysis, 
		we first introduce the weighting matrix
		$\mathbf Q=\operatorname{diag}(\frac{r^4}{\sin^2\theta},r^2)$
		and the covariance matrix
		\begin{equation}
			\mathbf\Sigma(\mathbf m)
			= \mathbf\Phi_1(\mathbf m)
			- \mathbf\Phi_2(\mathbf m)\mathbf\Phi_2^{\mathrm T}(\mathbf m)
			= \operatorname{Cov}\left(
			\begin{bmatrix}
				X+\gamma_1X^2\\
				\gamma_2X^2
			\end{bmatrix}
			\right),
			\label{eq:Sigma_random_variable_form}
		\end{equation} with $\gamma_1=\frac{\cos\theta}{r}$,
		$\gamma_2=\frac{\sin^2\theta}{2r^2}$,
		{\small
		\setlength{\arraycolsep}{3pt}
		$\mathbf\Phi_1(\mathbf m)=
		\begin{bmatrix}
			m_2+2\gamma_1m_3+\gamma_1^2m_4
			& \gamma_2(m_3+\gamma_1m_4)\\
			\gamma_2(m_3+\gamma_1m_4)
			& \gamma_2^2m_4
		\end{bmatrix}$},
		and
		{\small
		\setlength{\arraycolsep}{3pt}
		$\mathbf\Phi_2(\mathbf m)=
		\begin{bmatrix}
			m_1+\gamma_1m_2\\
			\gamma_2m_2
		\end{bmatrix}$}.
		Then, $f(\mathbf m)$ can be written as the weighted trace
		of the inverse covariance matrix
		\begin{equation}
			f(\mathbf m)
			= \operatorname{tr}\left(
			\mathbf Q\mathbf\Sigma^{-1}(\mathbf m)
			\right).
		\end{equation}
	 Consequently, we have the following proposition that claims the convexity of problem $\mathrm{(P_{2.1}')}$.
    \begin{proposition}
    	\label{prop_f_convex}
    	Problem $\mathrm{(P_{2.1}')}$ is a convex optimization problem.
    \end{proposition}
    \begin{IEEEproof}
    	Please see Appendix~A.
    \end{IEEEproof}
    To facilitate analysis, we introduce an auxiliary symmetric matrix $\mathbf W\in\mathbb S^2$ and define
	\begin{equation}
		\mathbf{H}(\mathbf m, \mathbf W)
		=
		\begin{bmatrix}
			\mathbf\Phi_1(\mathbf m)
			&
			\mathbf\Phi_2(\mathbf m)
			&
			\mathbf I_2
			\\
			\mathbf\Phi_2^\mathrm{T}(\mathbf m)
			&
			1
			&
			\mathbf 0_{1\times2}
			\\
			\mathbf I_2
			&
			\mathbf 0_{2\times1}
			&
			\mathbf W
		\end{bmatrix}.
		\label{eq:lifted_information_LMI}
	\end{equation}
Using this matrix, we obtain the following equivalent SDP formulation.
	\begin{proposition}
		The convex problem $\mathrm{(P_{2.1}')}$ can be equivalently reformulated
		as the following SDP:
		\begin{subequations}
			\begin{alignat}{2}
				\mathrm{(P_{2.1}'')}\quad
				& \operatorname*{minimize}_{
					\mathbf{m}\in\mathbb{R}^5, \mathbf{W}}
				&\quad&
			    \operatorname{tr}(\mathbf Q\mathbf W)
				\\
				& \hspace{0.1cm} \mathrm{s.t.}
				&&
				\mathbf{H}(\mathbf m, \mathbf W)\succeq\mathbf 0,
				\label{eq. p2-prime-cons-distri-1}
				\\
				&&&\eqref{eq. p2-cons-distri-1}~\mathrm{and}~\eqref{eq. p2-cons-distri-2}. \nonumber
			\end{alignat}
		\end{subequations}
	\end{proposition}
	\begin{IEEEproof}
		Please see Appendix~B.
	\end{IEEEproof}
	Since problem $\mathrm{(P_{2.1}'')}$ is a convex SDP, it can be solved using
	standard interior-point solvers
	through CVX \cite{cvx}. The resulting moment vector $\mathbf m^\star$ is globally
	optimal for problem $\mathrm{(P_{2.1})}$.
	\subsection{Stage II: Flat Extension and Atomic Extraction}
	In stage II, we aim to construct a feasible antenna position distribution $w(\cdot)$ based on the degree-$4$ truncated moment sequence $\mathbf m^\star$, which is a typical A-truncated $K$-moment problem (A-TKMP) in moment problem theory\footnote{Let $A=\{0,1,2,3,4\}$ index the moments up to degree $4$. Given a truncated moment sequence $\mathbf m|_A$, the A-TKMP seeks to determine whether $m$ admits a distribution supported on $K$ and, if so, how such a distribution can be constructed.} \cite{fialkow2016}.
	
	Generally, the degree-$4$ truncated moment sequence
	$\mathbf m^\star$
	obtained in Stage~I may admit multiple representing probability distributions supported on $[-a,a]$. To facilitate practical antenna deployment, we prefer a finitely atomic distribution, whose support points and weights specify the antenna locations and allocation proportions, respectively. The existence of such a distribution is guaranteed by \cite{NieFoCM,CurtoFialkow1998}. To construct it explicitly, we use the algorithm in \cite{NieFoCM} to search for a flat higher-order extension of $\mathbf m^\star$. Once the flatness condition is satisfied, we extract the corresponding support points and weights from the extended truncated moment sequence \cite{NieFoCM,CurtoFialkow1998}.	
    
    \subsubsection{Flat Extension}
    First, we define
    $\mathbf M_k(\mathbf z)
    = \big[z_{i+j}\big]_{0\le i,j\le k}$
    as the $k$-th order moment matrix, and
    $\mathbf L_k(\mathbf z)
    = \big[a^2z_{i+j}-z_{i+j+2}\big]_{0\le i,j\le k-1}$
    as the $k$-th order localizing matrix. Then, we introduce the flat extension theorem as shown below.
    
    \begin{theorem} \label{theorem:flat_extension}
    	Suppose that a degree-$2k$ truncated moment sequence
    	$\mathbf z=(z_0,\ldots,z_{2k})^{\mathrm T}$ satisfies
    	$
    		\mathbf M_k(\mathbf z)\succeq\mathbf 0
    	$, $
    		\mathbf L_k(\mathbf z)\succeq\mathbf 0,
    	$
    	and the flatness condition
    	\begin{equation}
    		\operatorname{rank}\mathbf M_k(\mathbf z)
    		=
    		\operatorname{rank}\mathbf M_{k-1}(\mathbf z)
    		=
    		\ell.
    		\label{eq:flat_condition}
    	\end{equation}
    	Then, $\mathbf z$ admits an $\ell$-atomic
    	distribution supported on [-a,a].
    \end{theorem}
	By Theorem~1, any flat extension of $\mathbf m^\star$ admits a finitely atomic representing distribution. We first check whether $\mathbf m^\star$ satisfies the flatness condition at order $k=2$. If so, the distribution can be extracted directly from $\mathbf m^\star$. Otherwise, following \cite{NieFoCM}, we solve the following convex optimization problem for successively higher orders $k=3,\ldots,k_{\mathrm{max}}$,\footnote{Theoretically, $k_{\mathrm{max}}$ can be limited up to 5 according to \cite{schmüdgen2017,NieFoCM}.} checking whether the resulting moment extension satisfies the flatness condition at each order:
	\begin{subequations}
		\begin{alignat}{2}
			\mathrm{(P_{2.2})} \quad
			& \operatorname*{minimize}_{
				\mathbf{z}} 
			&\quad&
			\sum_{j=0}^{6}R_j z_j, \label{eq:SDRk}
			\\
			& \hspace{0.1cm} \mathrm{s.t.}
			&& \hspace{-0.9cm} \mathbf L_k(\mathbf{z})\succeq\mathbf{0},\quad\mathbf M_k(\mathbf{z}) \succeq \mathbf{0},
			\label{eq. SDR-cons-distri-1}\\
			&&&\hspace{-0.9cm} z_i = m^\star_i,\quad i=0,1,2,3,4. 
		\end{alignat}
	\end{subequations}
	Here, $R_0,\ldots,R_6$ denote the coefficients of a fixed generic degree-six univariate sum of squares polynomial $R(x)=\sum_{j=0}^{6}R_jx^j$. In practice, $R(x)$ can be chosen in Gram form following
	\cite{NieFoCM}, i.e., 
	$
	R(x)=\mathbf v^{\mathrm T}(x)\mathbf G\mathbf v(x),
	$
	where $\mathbf v(x)=[1,x,x^2,x^3]^{\mathrm T}$ and $\mathbf G=\mathbf A^{\mathrm T}\mathbf A$, with the entries of $\mathbf A\in\mathbb R^{4\times4}$ independently drawn from the standard normal distribution. The resulting $R(x)$ is kept fixed as the relaxation order increases.
	Starting with $k=3$, we solve $(\mathrm P_{2.2})$ using CVX and denote an optimizer by $\mathbf z^{(k)}$. We then check whether $\mathbf z^{(k)}$ satisfies the flatness condition \eqref{eq:flat_condition}. If so, Theorem~1 guarantees that $\mathbf z^{(k)}$ admits an $\ell$-atomic representing probability distribution supported on $[-a,a]$. Its support points and weights are then extracted as described subsequently. Otherwise, we increase $k$ by one and repeat the optimization and flatness check.
	\subsubsection{Support and Weight Extraction}
	\label{subsec:extraction}
	Having obtained a flat extension $\mathbf z^\star=\mathbf z^{(k)}$, we construct the corresponding $\ell$-atomic representing probability distribution by recovering its support points via the univariate Prony method \cite{KunisLAA2016} and determining its weights from a Vandermonde system.
	
	For moment matrix $\mathbf M_k(\mathbf z^\star)$ satisfying the flatness condition \eqref{eq:flat_condition}, 
	its first $\ell$ columns
	$\mathrm{col}_0(\mathbf M_k(\mathbf z^\star)),\ldots,\mathrm{col}_{\ell-1}(\mathbf M_k(\mathbf z^\star))$
	form a basis for its column space. Consequently, the next column admits the unique representation
	\begin{equation}
		\mathrm{col}_{\ell}(\mathbf M_k(\mathbf z^\star))
		=
		\sum_{j=0}^{\ell-1}c_j\mathrm{col}_j(\mathbf M_k(\mathbf z^\star)),
		\label{eq: col_relation}
	\end{equation}
	where the coefficients $c_0,\ldots,c_{\ell-1}$ are uniquely determined by \eqref{eq: col_relation}.
	We then define the associated Prony polynomial 
	$
		\rho(\zeta)
		=
		\zeta^\ell-\sum_{j=0}^{\ell-1}c_j\zeta^j
	$. The column relation \eqref{eq: col_relation} implies that
	$\rho(\zeta)$ vanishes at every support point of the corresponding $\ell$-atomic distribution \cite{KunisLAA2016}. Since $\rho(\zeta)$ has degree $\ell$ and the distribution has $\ell$ distinct support points, these points are precisely its roots. Thus, the atomic locations $\xi_1^\star,\ldots,\xi_\ell^\star$ are recovered as the $\ell$ distinct real roots of $\rho(\zeta)$ in $[-a,a]$.
	
	
	Having recovered the support points $\xi_1^\star,\ldots,\xi_\ell^\star$, we determine the corresponding weights $\boldsymbol\omega=[\omega_1,\ldots,\omega_\ell]^{\mathrm T}$. Note that the $j$-th moment satisfies $z_j^\star=\sum_{i=1}^{\ell}\omega_i(\xi_i^\star)^j$. Solving the first $\ell$ moment equations, corresponding to $j=0,\ldots,\ell-1$, yields
    \begin{equation}
    	\boldsymbol\omega
    	=
    	\mathbf V^{-1}
    	[z_0^\star,\ldots,z_{\ell-1}^\star]^{\mathrm T},
    \end{equation}
    with the Vandermonde matrix
\begin{equation}
	\mathbf V
	=
	\left[
	\begin{smallmatrix}
		1&1&\ldots&1\\
		\xi_1^\star&\xi_2^\star&\ldots&\xi_\ell^\star\\
		\vdots&\vdots&\ddots&\vdots\\
		(\xi_1^\star)^{\ell-1}&
		(\xi_2^\star)^{\ell-1}&
		\ldots&
		(\xi_\ell^\star)^{\ell-1}
	\end{smallmatrix}
	\right]
	\in\mathbb R^{\ell\times\ell}.
	\label{eq:Vandermonde_definition}
\end{equation} Here the matrix $\mathbf V$ is nonsingular since the support points are distinct.
	Consequently, the resulting $\ell$-atomic antenna position distribution
	is reconstructed as
	$
		w^\star(\cdot)
		=
		\sum_{i=1}^{\ell}
		\omega_i \delta_{\xi_i^\star}(\cdot).
	$
	\subsubsection{Practical Implementation}
Having obtained the optimal $\ell$-atomic antenna position distribution,
we next construct a practical discrete placement with a given
number of antennas $N$, subject to the minimum inter-element constraint \eqref{eq. cons-distri-1}.
Specifically, the weight $\omega_i$ associated with the
$i$-th support point is converted into a number of antennas as
$N_i=\operatorname{round}\!\left(N\omega_i\right),$ for $i=1,\ldots,\ell-1$, and the antenna counts at the last support point are determined by $
N_\ell
=
N-\sum_{i=1}^{\ell-1}N_i.
$
Each reconstructed point is thereby realized by a local cluster
of $N_i$ antennas. Within the $i$-th cluster, the antennas are
uniformly placed with spacing $\frac{\lambda}{2}$ within the given aperture.
	\subsection{Computational Complexity Analysis}
	In this subsection, we analyze the computational complexity of the proposed two-stage
	algorithm.
	In Stage I, problem $(\mathrm{P}_{2.1}'')$ has a fixed number of
	variables and fixed-size semidefinite constraints. Its computational cost
	is therefore $\mathcal{O}(\log(1/\epsilon))$, with $\epsilon \in (0,1)$ denoting the accuracy of the interior-point method \cite{Bental}.
	In Stage II, the $k$-order extension problem $(\mathrm{P}_{2.2})$
	contains $2k-4$ scalar variables and two semidefinite
	blocks of sizes $k+1$ and $k$ respectively. Following \cite{Bental}, 
	each interior-point iteration in $(\mathrm{P}_{2.2})$ costs $\mathcal{O}(k^4)$,
	and iteration complexity is $\mathcal{O}(\sqrt{k}\log(1/\epsilon))$,
	yielding $\mathcal{O}(k^{9/2}\log(1/\epsilon))$ per extension.
	With the extension orders $k_{\mathrm{max}}$ bounded by $5$, at most
	three SDPs are solved, yielding the total
	cost of the extension SDPs $\mathcal{O}(\log(1/\epsilon))$.
	The flatness checks, support and weight extraction require
	only fixed-size linear algebra and polynomial root computations,
	which contribute constant costs.
	Finally, 
	the overall complexity of the two-stage algorithm is therefore
	$\mathcal{O}(\log(1/\epsilon))$, which is independent of the given number of antennas $N$. Thus, the computational cost remains constant with respect to $N$,
	even for large antenna arrays.

	\section{Simulation Results}
     In this section, we evaluate the proposed two-stage algorithm and compare the proposed antenna placement design strategy against conventional array placements and the exhaustive search baseline. Throughout the simulations, we set the carrier frequency
     $f_0=28~\mathrm{GHz}$ \cite{liuarxiv2609}, the number of snapshots
     $T=1024$, a half-aperture of
     $a=50\lambda$, $N=45$ antennas, and $\mathrm{SNR} = \frac{|\alpha|^2P}{\sigma^2}$ to
     $10~\mathrm{dB}$ \cite{LiuTWC0426}, unless stated otherwise.
     \subsection{Constructed Antenna Position Distribution and Practical Placement Analysis}
     
     First, we evaluate the constructed atomic distributions obtained in the proposed two-stage algorithm.
     \begin{table}[t]
     	\centering
     	\caption{Constructed antenna position distributions at different target locations.}
     	\label{tab:atomic_measures}
     	\footnotesize
     	\renewcommand{\arraystretch}{1.05}
     	\setlength{\tabcolsep}{2pt}
     	\begin{tabular}{@{}c|cccc@{}}
     		\toprule
     		\multirow{2}{*}{Quantity}
     		& \multicolumn{4}{c}{Target location $(p_x,p_y)$ in meters} \\
     		\cmidrule(lr){2-5}
     		& $(0,20)$
     		& $(12,12)$
     		& $(3,1)$
     		& $(1,0.3)$ \\
     		\midrule
     		
     		$\ell$
     		& $3$ & $3$ & $3$ & $3$ \\
     		
     		$(\xi_1^\star,\xi_2^\star,\xi_3^\star)$
     		& \multicolumn{4}{c}{$(-a,\,0,\,a)$} \\
     		
     		$\omega_1^\star$
     		& $0.250$
     		& $0.256$
     		& $0.290$
     		& $0.372$ \\
     		
     		$\omega_2^\star$
     		& $0.500$
     		& $0.500$
     		& $0.500$
     		& $0.499$ \\
     		
     		$\omega_3^\star$
     		& $0.250$
     		& $0.244$
     		& $0.210$
     		& $0.129$ \\
     		\bottomrule
     	\end{tabular}
     	\vspace{-5mm}
     \end{table}
     As shown in Table~\ref{tab:atomic_measures}, for each considered sensing target location, the extraction procedure yields an antenna position distribution supported at the center and two edges of the aperture, i.e., $(\xi_1^\star,\xi_2^\star,\xi_3^\star)=(-a,0,a)$.
     Let $(\omega_1,\omega_2,\omega_3)$ denote the corresponding weights.
     As observed, the two edge weights are equal for a broadside sensing target, yielding a centro-symmetric antenna position distribution.
    Additionally, the edge weights become increasingly asymmetric as the target moves toward the positive-$x$ side of the array, with $\omega_1$ increasing and $\omega_3$ decreasing.
     \begin{figure}[t]
     	\centering
     	\includegraphics[width=0.8\linewidth, height=0.8\linewidth, keepaspectratio]{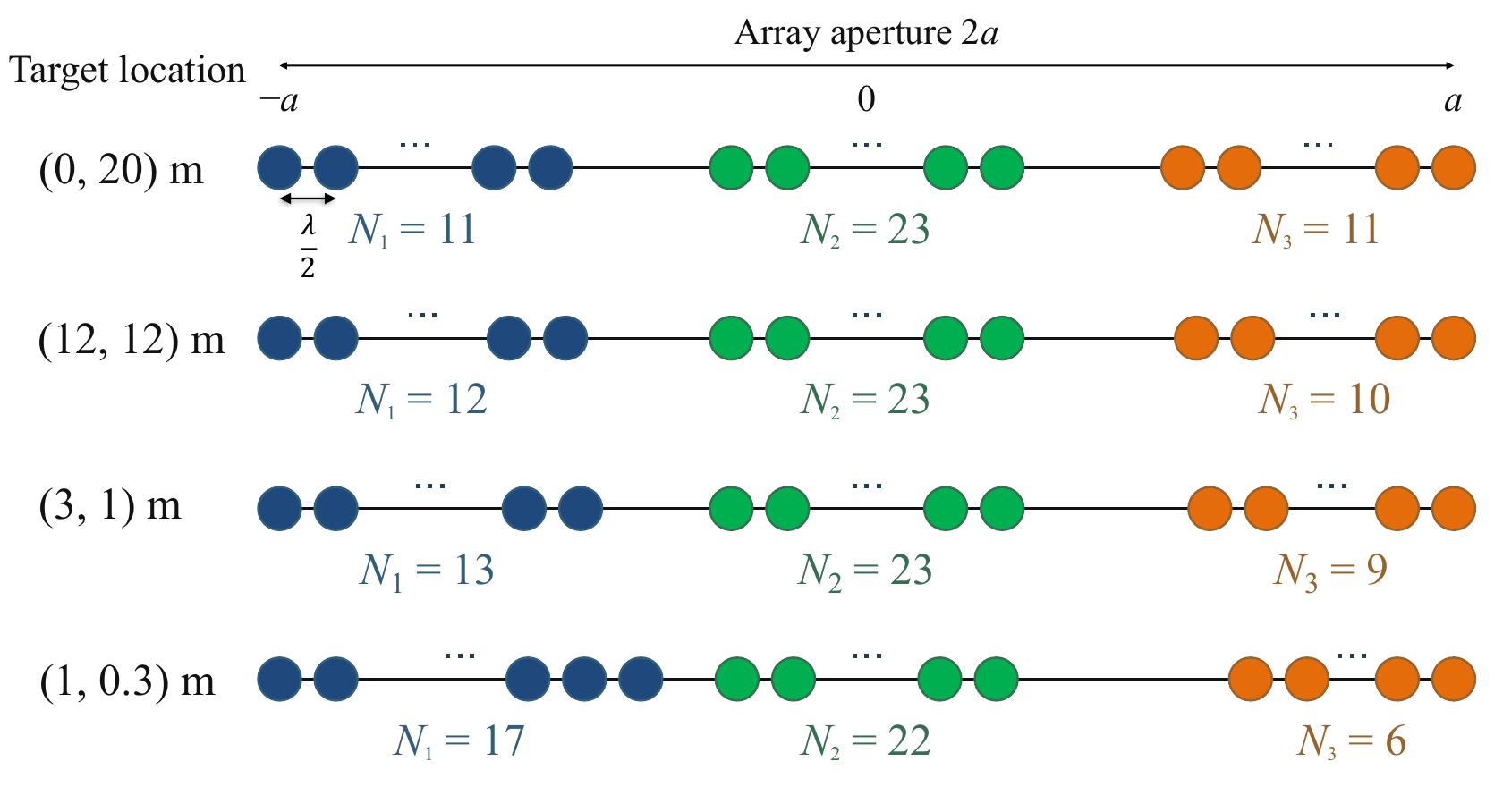}
     	\vspace{-2mm}
     	\caption{Practical antenna position placements at different target locations.}
     	\vspace{-2mm}
     	\label{fig:deployment}
     \end{figure}
     
     Fig.~\ref{fig:deployment} illustrates the corresponding practical
     placements with a minimum spacing of $\lambda/2$.
     Consistent with the observations in Table~\ref{tab:atomic_measures}, more antennas are
     allocated to the left cluster as the target location
     departs further from broadside toward the positive-$x$ axis
     in the considered cases. This can be intuitively explained by a fact that allocating more antennas to the left cluster
     helps compensate for the angle-range coupling induced
     by the off-broadside target location, thereby providing greater position information and thus improving
     localization accuracy.

     \subsection{Antenna Placement Analysis}
\begin{figure}[t]
   	\centering
   	\includegraphics[width=0.48\linewidth, height=0.48\linewidth, keepaspectratio]{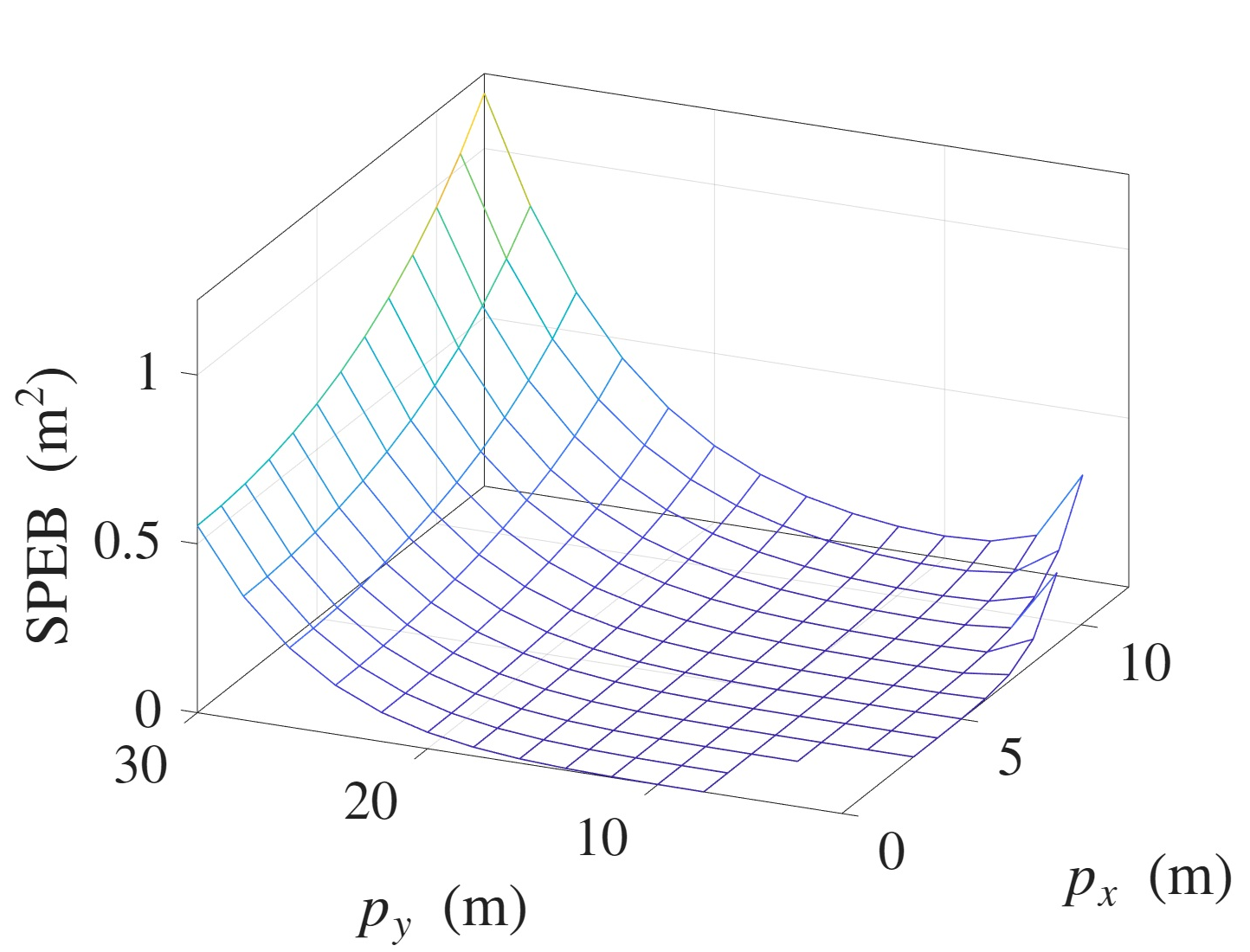}
   	\vspace{-2mm}
   	\caption{SPEB versus the coordinates of sensing target.}
   	\label{fig:SPEB_surface}
   	\vspace{-5mm}
     \end{figure}

   Next, we evaluate how target coordinates affect sensing performance. Fig.~\ref{fig:SPEB_surface} shows the SPEB achieved by the placement optimized for each target location. Over the near-field region $\Gamma$, the SPEB increases with either $p_x$ or $p_y$ when the other coordinate is fixed. This trend is consistent with the stronger path loss and reduced phase sensitivity to range at larger sensing target distances, as reflected in \eqref{eq.FIM-monostatic}.
     
     Then, we benchmark the proposed placement policy against the following three conventional array geometries and the exhaustive search baseline:
     \begin{itemize}
     	\item \textbf{ULA ($\lambda/2$ spacing) \cite{Balanis}:}
     	The antennas are uniformly spaced by $\lambda/2$ and centered within
     	the available aperture.
     	
     	\item \textbf{Sparse ULA \cite{ShiTSP2021}:}
     	The antennas are uniformly distributed over the entire aperture.
     	
     	\item \textbf{Two-edge array \cite{MaTWC0824}:}
     	The antennas are equally allocated to two clusters at
     	aperture ends with inter-spacing $\lambda/2$.
     	
     	\item \textbf{Exhaustive search:}
     	The antenna positions are numerically optimized by exhaustive search.
     \end{itemize}
     
     \begin{figure}[t]
     	\centering
     	\subfloat[$(p_x,p_y)=(12,12)\,\mathrm{m}$
     	\label{fig:speb-G1-SNR}]{%
     		\includegraphics[width=0.48\linewidth]{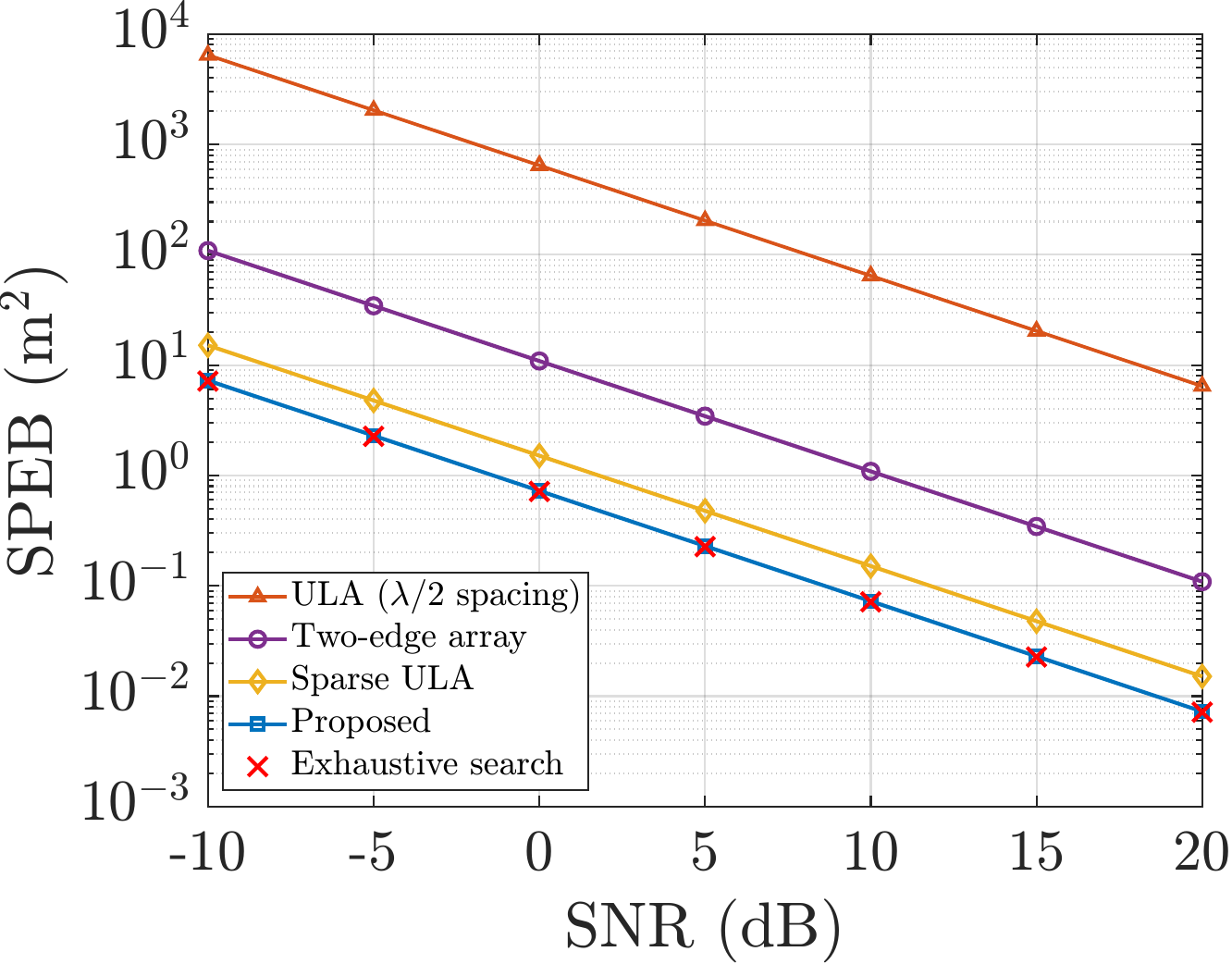}%
     	}
     	\hfill
     	\subfloat[$(p_x,p_y)=(3,1)\,\mathrm{m}$
     	\label{fig:speb-G2-N}]{%
     		\includegraphics[width=0.48\linewidth]{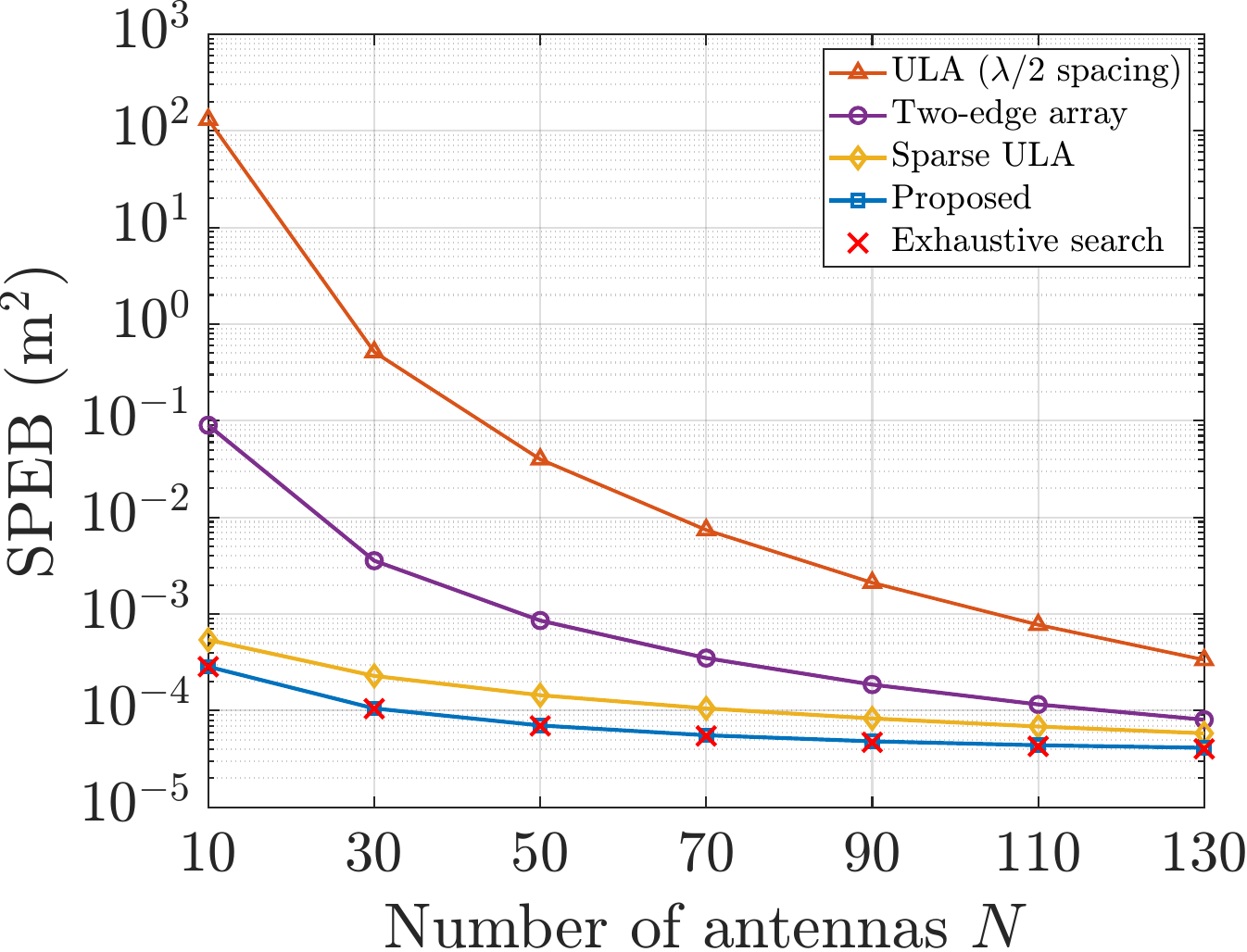}%
     	}
     	\caption{SPEB versus (a) SNR and
     		(b) number of receive antennas.}
     	\label{fig:speb-G-SNR}
     	\vspace{-5mm}
     \end{figure}
     Fig.~\ref{fig:speb-G-SNR} shows the SPEB versus
     SNR and the number of antennas $N$ at two different target locations, respectively.
     In Fig.~\ref{fig:speb-G-SNR}(a), the SPEB decreases inversely
     with the SNR, consistent with
     $\kappa\propto\mathrm{SNR}^{-1}$ in \eqref{eq.FIM-monostatic}.
     In Fig.~\ref{fig:speb-G-SNR}(b), increasing $N$ improves
     estimation performance by providing additional spatial
     observations, while the reduction of SPEB becomes more gradual at larger $N$. 
     The improvement is particularly pronounced for the ULA
     and two-edge placement. This is because increasing $N$ enlarges the ULA aperture
     and extends the edge clusters toward the array center
     in the two-edge placement, providing more diverse target location information in
     phase responses that enables better localization accuracy. In all considered cases, the proposed placement outperforms the
     conventional schemes and closely matches the
     exhaustive search baseline at much lower computational cost, demonstrating the effectiveness of the proposed design.
    \section{Conclusion}
    In this paper, we investigate an MA-enabled antenna placement problem for near-field sensing. We first show that the SPEB depends on the antenna position distribution only through its degree-4 truncated moments, enabling an equivalent SDP that can be solved by standard convex solvers. Then we recover an optimal antenna position distribution with limited clusters, and construct a practical deployment strategy. Numerical results show that the proposed placement design consistently outperforms conventional array geometries and closely matches the performance of exhaustive search across different SNRs and antenna counts, at substantially lower computational cost.
	\appendices
	\section{Proof of Proposition~1}
		Let $\mathbf m_1$ and $\mathbf m_2$ be feasible moment
		vectors
		and define
		$\mathbf m_\tau=\tau\mathbf m_1+(1-\tau)\mathbf m_2$
		for $\tau\in[0,1]$.
		Since $\mathbf\Phi_1$ and $\mathbf\Phi_2$ are affine in $\mathbf{m}$, it follows that
		\begin{align}
			\mathbf\Sigma(\mathbf m_\tau)
			&=
			\tau\mathbf\Sigma(\mathbf m_1)
			+(1-\tau)\mathbf\Sigma(\mathbf m_2)
			+\tau(1-\tau)\mathbf P\mathbf P^{\mathrm T}
			\nonumber\\
			&\succeq
			\tau\mathbf\Sigma(\mathbf m_1)
			+(1-\tau)\mathbf\Sigma(\mathbf m_2), \label{eq:Sigma_matrix_concave}
		\end{align}
		where
		$\mathbf P=\mathbf\Phi_2(\mathbf m_1)
		-\mathbf\Phi_2(\mathbf m_2)$.
		Thus, $\mathbf\Sigma(\mathbf{m})$ is matrix concave in $\mathbf{m}$ and function
		$
		\operatorname{tr}
		\left(
		\mathbf Q\mathbf W^{-1}
		\right)
		$
		is convex and monotonically decreasing in Loewner
		order. Consequently, by
		\eqref{eq:Sigma_matrix_concave},
		$
			f\!\left(\mathbf m_{\tau}\right)
			\le
			\tau f\!\left(\mathbf m_1\right)
			+(1-\tau)f\!\left(\mathbf m_2\right).
		$
		Thus, $f(\mathbf m)$ is convex and the moment and localizing matrices are affine in $\mathbf m$ in \eqref{eq. p2-cons-distri-1} hence are linear matrix inequalities. Therefore, the problem $(\mathrm{P}_{2.1}')$ is convex.
	\section{Proof of Proposition~2}
		Applying the Schur complement to the scalar block $1$
		in $\mathbf H(\mathbf m,\mathbf W)$ in
		\eqref{eq:lifted_information_LMI} shows that
		$\mathbf H(\mathbf m,\mathbf W)\succeq\mathbf 0$ is equivalent to 
		$[\boldsymbol{\Sigma}(\mathbf m),\,\mathbf I_2;
		\mathbf I_2,\,\mathbf W]\succeq\mathbf 0$.
		The identity off-diagonal block implies
		$\mathbf\Sigma(\mathbf m)\succ\mathbf 0$, since any nonzero vector in its null space would
		contradict the PSD of the block matrix. Taking the Schur complement of
		$\mathbf\Sigma(\mathbf m)\succ\mathbf 0$
		in this block matrix gives
		$
			\mathbf W-\mathbf\Sigma(\mathbf m)^{-1}
			\succeq\mathbf 0.
		$
		Hence, $\mathbf H(\mathbf m,\mathbf W)\succeq\mathbf 0$
		is equivalent to
		$\mathbf\Sigma(\mathbf m)\succ\mathbf 0$
		and $\mathbf W\succeq\mathbf\Sigma(\mathbf m)^{-1}$.
		Since $\mathbf Q\succeq\mathbf 0$, it follows that
		$
			\operatorname{tr}(\mathbf Q\mathbf W)
			\geq
			\operatorname{tr}\!\left(
			\mathbf Q\mathbf\Sigma(\mathbf m)^{-1}
			\right)
			=f(\mathbf m),
		$
		with equality attained at
		$\mathbf W=\mathbf\Sigma(\mathbf m)^{-1}$
		for every feasible $\mathbf m$.
		Thus, $(\mathrm{P}_{2.1}'')$ exactly reformulates
		$(\mathrm{P}_{2.1}')$.
		Its objective $\operatorname{tr}(\mathbf Q\mathbf W)$
		is linear, and all constraints are linear matrix
		inequalities or affine equalities.
		Therefore, $(\mathrm{P}_{2.1}'')$ is a convex SDP.
	\bibliographystyle{IEEEtran}
	\bibliography{references}   
	
\end{document}